\documentclass[article]{elsarticle}

\usepackage{mathtools}
\usepackage{amssymb}
\usepackage{lineno,hyperref}
\usepackage{graphicx}
\usepackage{subcaption}
\usepackage{float}
\modulolinenumbers[5]
\journal{Macromolecular Theory and Simulations}
\modulolinenumbers[5]
\usepackage{textcomp}
\usepackage{amssymb}  
\usepackage{bm}       
\usepackage{booktabs} 
\usepackage{dcolumn}  
\usepackage{multirow} 
\usepackage{graphicx} 
\usepackage[printonlyused]{acronym}
\usepackage{mathtools}
\usepackage{lineno,hyperref}
\usepackage{subcaption}
\usepackage{float}
\usepackage{ragged2e}
\usepackage{amsmath}

\usepackage[usenames, dvipsnames]{color}
\graphicspath{{fig2/}} 
\begin{document}
	
	\begin{frontmatter}
		
		\title{CFD Investigation of Thermal performance of Aluminum oxide nanofluid in channel }
		
		\author{Mehdi Jafari}
		\author{Mohammad Khalili\corref{cor1}}
		\cortext[cor1]{Corresponding author}
		
		\address{Khomeinishahr Branch, Islamic Azad University, Khomeinishahr, Iran}

		\begin{abstract}
		The present paper is a numerical study of heat transfer and pressure drop of two nano-fluids including water as base fluid with Al${}_{2}$O${}_{3}$ nano-particles inside a square channel having a cylinder inside, with and without fin under constant heat flux condition by using two-phase Euler-Lagrange approach. In this paper, numerical investigation has been done for various combinations of base fluid, nano-particle size, and concentration through a straight cylinder. Simulation has been performed in a laminar flow regime using finite volume method. Besides, the thermal boundary condition of constant uniform heat flux on the cylinder wall was applied. The results show that the increase of Reynolds number and nano-particles volume concentration have considerable effects on heat transfer coefficient enhancement. For nano-particles, the heat transfer coefficient decreases when nano-particles diameter increases. The passive way used in this study leads to higher pressure drops. For all fluids under consideration, pressure drop escalated with Reynolds number. Adding nano-particles to the base fluid leads to rise in pressure drop and this effect is more intensive for higher concentrations. Regardless of nano-particles type and their volume concentration, the skin friction coefficient increases with a rise in Reynolds number.
		\end{abstract}
		
		
	\end{frontmatter}

	\section{Introduction}
	Lots of changes have happened in the field of enhanced heat transfer. Offset strip fins are designed to enhance heat transfer by enlarging surface. Some reviews of experimental studies on the effects of different surface geometries on the flow and heat transfer performance of offset strip fins are addressed in \cite{alipour2017influence,darzi2016two,hasan2014investigation}.
	
	\noindent Suspensions of nano-particles also can increase the viscosity and decrease the specific heat, which means that the improvement in thermal conductivity of nano-fluids may be counteracted by the negative effects of viscosity and specific heat. A wide variety of processes involving heat and energy, have low efficiency due to low thermal conductivity of working fluid such as water, ethylene glycol, and engine oil. An advanced method proposed to enhance heat transfer characteristic of fluid is dispersion of ultrafine solid particles in a base fluid known as nano-fluids. There are considerable researches on the superior heat transfer properties of nano-fluids especially on thermal conductivity and convective heat transfer \cite{webb2005enhanced,sohel2014experimental}.
	
	\noindent In the recent decade, numerous investigations are performed by using two-phase method in which mass, momentum, and energy conservation equations are solved for each phase individually or momentum and energy conservation equations coupled with mass conservation equation are solved for each phase. Ali Akbari et al. \cite{akbari2016modified} numerically studied the heat transfer of CuO nano-fluid through a micro-tube using two phase approach. They showed that increasing nano-particles volume concentration leads to reduction in Nusselt number and friction factor. Moshizi et al. \cite{moshizi2014two} Investigated heat transfer and pressure drop properties of Al${}_{2}$O${}_{3}$--H${}_{2}$O nano-fluid through a tube under constant heat flux condition. Results indicate that slip velocity at the tube walls improves coefficient of heat transfer and enhances the ratio of pressure gradient. Heat transfer under nano-fluids flow with non-Newtonian base fluid inside a microchannel was studied by Esmaeilnejad et al. \cite{esmaeilnejad2014numerical} applying two-phase mixture model. Results obtained illustrated impressive heat transfer enhancement of non-Newtonian fluids where nano-particles are distributed. By rising volume fraction, more heat transfer enhancement would be achieved.
	
	\noindent Due to a wide range of practical applications in heat transfer systems, many investigations on nano-fluids reported that nano-fluids have desirable properties and behaviors such as enhanced wetting and spreading \cite{chengara2004spreading}, increased heat transfer in forced convection \cite{mansour2007effect,nguyen2007heat,palm2006heat}.
	
	\noindent Shuai and Chang \cite{shi2011numerical} used a new type of non-continuous finned tubes that enhances heat transfer area. They found that the use of finite volume method is very helpful and important in design of three-dimensional pin fin tubes.
	
	\noindent Mir et al. \cite{mir2004numerical} studied laminar forced convection in a finned annulus numerically. The simulation corresponded to a thermal boundary condition of uniform heat input per unit axial length with peripherally uniform temperature at any cross section. In the solution domain, heat transfer and fluid flow were investigated for different ratios of inner and outer pipes radius, fins height and number of fins. The results showed good agreement with other publications.
	
	\noindent K. Goudarzi and H. Jamali \cite{goudarzi2017heat} used aluminum oxide in ethylene glycol as a nano-fluid so as to enhance heat transfer properties in a car radiator with wire coil inserts. The results showed enhancement in heat transfer rates up to 9\% if using coils. Furthermore, the simultaneous use of coils with nano-fluids caused by the better thermal performance enhancement than coils alone.
	
	\noindent Ding et al. \cite{ding2009parametric} presented a numerical study using TiO${}_{2}$-water as a nano-fluid. The flow was laminar under a constant wall heat flux prevailed. The study was carried out for both heat transfer coefficient and wall shear stress. By increasing nano-particles volume concentration, the shear stress increased.
	
	\noindent Sohel et al. \cite{sohel2014experimental} experimentally showed that by increasing volumetric concentration of Al${}_{2}$O${}_{3}$-water nano-fluid, thermal effectiveness increased at all flow rates. However, they found that thermal effectiveness was not necessarily increased with the increase of flow rate. They found 18\% convective heat transfer coefficient enhancement by using 0.25\% concentrated Al${}_{2}$O${}_{3}$-water nano-fluid as compared to distilled water.
	
	\noindent Ho et al. \cite{ho2010experimental} assessed Al${}_{2}$O${}_{3}$-water nano-fluid forced convection heat transfer. They observed 1\% volumetric concentration nano-fluid was more efficient than 2\% due to more variation occurrence in dynamic viscosity with temperature. Using 1\% volume concentrated Al${}_{2}$O${}_{3}$-water nano-fluid, 70\% enhancement was found in convective heat transfer coefficient.
	
	\noindent S. Akbarzadeh et al. \cite{akbarzadeh2014experimental} performed a study about the influence of particles volume fraction and temperature on the thermal conductivity and viscosity\textbf{ }of nano-fluids and obtained new correlations. The results showed that viscosity of nano-fluids increases if any rise occurs in particles volume concentration. The thermal conductivity exhibited a nonlinear growth with the particle volume fraction. Furthermore, in high temperatures, thermal conductivity experienced an improvement.
	
	\noindent Cong et al. \cite{nguyen2007heat} studied heat transfer enhancement and behavior of Al${}_{2}$O${}_{3}$ nano-particles for use in cooling systems such as microprocessors or electronic components. Results showed a very good enhancement in heat transfer properties. For example, for a special particle volume concentration of 6.8\%, the heat transfer coefficient increased almost 40\% in comparison with the base fluid.
	
	\noindent Zhang et al. \cite{zhang2017experimental} investigated heat transfer enhancement making use of micro-fin structures and nano-fluids. Data obtained presented that by increasing number of fins, Nusselt number and friction factor increase. In addition, using TiO${}_{2}$-water nano-fluid prospers heat transfer process, but causes the pressure to drop.
	
	\noindent The literature shows the importance of heat transfer enhancement using nano-fluids and fins each individually or combined together. Looking into conducted researches, it seems that there are very limited studies on non-circular channels for nano-fluids and heat transfer. Also, the effects of longitudinal fin and its shape in the channel on heat transfer have not been investigated so far. The objective of this paper is to study heat transfer and pressure drop of two nano-fluids including water as the base fluid with Al${}_{2}$O${}_{3}$ nano-particles inside a square channel having a cylinder inside, with and without fin under constant heat flux condition.
	
	\section{Physical Model}
	\subsection{Geometry}
	\noindent As mentioned earlier, the purpose of this paper is to study a nano-fluid flow in a channel having an inner cylinder. For this reason, Finite volume method has been used. In the current research, a square channel with a rigid cylinder located in its center and a flat and wavy separating sheet are used. In Fig. 1, a schematic of channel geometry is exhibited in non-real dimensions. Channel dimensions are shown in Fig. 1 and presented in Table 1 in which L, A and B are length, width, and height of the channel, respectively and D is rigid cylinder diameter.
	\subsection{Governing Equations}
	
\noindent Numerous studies have been done in order to model two-phase flows in which acceptable results are achieved  \cite{khodabandeh2018performance,KHODABANDEH201743,aminfar20113d}. Two-phase Eulerian-Lagrangian model is considered as a continuous medium, while fluid and nano-particles phases are considered as discrete mediums. Thus, The Navier-Stokes equations are required to be solved for the fluid. Buongiorno \cite{buongiorno2008nanofluids} concluded that among diffusion mechanisms having an effect on nano-particles, thermophoretic and Brownian forces are amongst the most important ones. For this reason, effects of thermophoretic and Brownian forces are expressed as source terms while applying momentum equation. Therefore, mass, momentum, and energy conservation equations for the liquid phase being exposed at a steady state flow and under the condition of temperature-dependent physical properties are written as

\noindent Mass conservation equation is written as \cite{khodabandeh2016effects}:

\begin{equation}
	\nabla \bullet ( \rho_f V_f)=0
\end{equation}

\noindent Momentum conservation equation is written as \cite{khodabandeh2016effects}:

\begin{equation}
	\mathrm{\nabla }\mathrm{\bullet }\left({\mathrm{\rho }}_{\mathrm{f}}{\mathrm{V}}_{\mathrm{f}}{\mathrm{V}}_{\mathrm{f}}\right)\mathrm{=-}\mathrm{\nabla }\mathrm{P+}\mathrm{\nabla }\mathrm{\bullet }\left({\mathrm{\mu }}_{\mathrm{f}}\mathrm{\nabla }{\mathrm{V}}_{\mathrm{f}}\right)\mathrm{+}\mathrm{\rho }\left({\mathrm{F}}_{\mathrm{B}}\mathrm{+}{\mathrm{F}}_{\mathrm{D}}\mathrm{+}{\mathrm{F}}_{\mathrm{L}}\mathrm{+}{\mathrm{F}}_{\mathrm{V}}\mathrm{+}{\mathrm{F}}_{\mathrm{G}}\mathrm{+}{\mathrm{F}}_{\mathrm{P}}\right)
\end{equation}

\noindent In which $F_B$, $F_D$, $F_L$, $F_V$, $F_G$ and $F_P$ are Brownian force, drag force, Saffman lift force, virtual mass force, gravitational force and gradient pressure force applied on a particle, respectively.  The Brownian force is applied on fine particles spread through the fluid. Collisions among particles and fluid molecules affect the dispersion of the particles when particles are fine enough and micron-sized. The influence of Brownian force is appropriated in an additional forces term.

\noindent Using the process of Gaussian spectrum with the intensity spectrum of $S_{n,ij}$, the Brownian force components can be modeled. Based on this fact, one can write \cite{li1992dispersion}

\begin{equation}
	S_{n,ij}=S_0{\delta }_{ij}
\end{equation}

In which ${\delta }_{ij}$ is introduced as the Kronecker delta function possessing the following form.

\begin{equation}
	{\mathrm{S}}_0\mathrm{=}\frac{\mathrm{21}\mathrm{\nu }{\mathrm{k}}_{\mathrm{B}}\mathrm{T}}{{\mathrm{\pi }}^{\mathrm{2}}\mathrm{\rho }{\mathrm{d}}^{\mathrm{5}}_{\mathrm{P}}{\left(\frac{{\mathrm{\rho }}_{\mathrm{P}}}{\mathrm{\rho }}\right)}^{\mathrm{2}}{\mathrm{C}}_{\mathrm{c}}}
\end{equation}

Where $T$ is the absolute fluid temperature, $\nu $ is kinematic viscosity and $k_B$ is Stefan-Boltzmann constant. The magnitude of the Brownian force components can be computed by

\begin{equation}
F_{B_i}\mathrm={\zeta }_i\sqrt{\frac{\pi S_0}{\mathrm{\Delta }t}}
\end{equation}                                                                                                                                                  
Where ${\zeta }_i$ is the Gaussian random numbers with the variance of non-unity and the mean value of zero.

\noindent The drag force is calculated from the following equation.

\begin{equation}
{\mathrm{F}}_{\mathrm{D}}\mathrm{=}\frac{\mathrm{18}\mathrm{\mu }}{{\mathrm{d}}^{\mathrm{5}}_{\mathrm{P}}{\mathrm{\rho }}_{\mathrm{P}}{\mathrm{C}}_{\mathrm{c}}}\left({\mathrm{V}}_{\mathrm{F}}\mathrm{-}{\mathrm{V}}_{\mathrm{P}}\right)   
\end{equation}

Where ${\rho }_P$ is density of the nano-particles, $C_c$ is the Cunningham correction factor that is computed by Eq. (7), $V_F$ and$\ V_P$ are the speed of continuous phase and particles, respectively.

\begin{equation}
{\mathrm{C}}_{\mathrm{c}}\mathrm{=1+}\frac{\mathrm{2}\mathrm{\lambda }}{{\mathrm{d}}_{\mathrm{P}}}\left(\mathrm{1}.\mathrm{257+0}.\mathrm{4}{\mathrm{e}}^{\mathrm{-}\frac{\mathrm{1}.\mathrm{1}{\mathrm{d}}_{\mathrm{P}}}{\mathrm{2}\mathrm{\lambda }}}\right) 
\end{equation}

Where $\lambda $ is defined as molecular free path. The gravitational force and gradient pressure force are acquired through Eq. (8) and (9), respectively.
\begin{equation}
{\mathrm{F}}_{\mathrm{G}}\mathrm{=}\frac{\mathrm{g}\left({\mathrm{\rho }}_{\mathrm{P}}\mathrm{-}{\mathrm{\rho }}_{\mathrm{f}}\right)}{{\mathrm{\rho }}_{\mathrm{P}}}                  
\end{equation}

\begin{equation}
{\mathrm{F}}_{\mathrm{P}}\mathrm{=}\left(\frac{{\mathrm{\rho }}_{\mathrm{f}}}{{\mathrm{\rho }}_{\mathrm{P}}}\right){\mathrm{V}}_{\mathrm{P}}\mathrm{\bullet }\mathrm{\nabla }{\mathrm{V}}_{\mathrm{f}} 
\end{equation}

The force needed to accelerate the fluid around the particles is called the virtual mass force which is stated as follows (Ranz- Marshal \cite{ranz1952evaporation}):

\begin{equation}
{\mathrm{F}}_{\mathrm{V}}\mathrm{=}\frac{\mathrm{1}}{\mathrm{2}}\frac{{\mathrm{\rho }}_{\mathrm{f}}}{{\mathrm{\rho }}_{\mathrm{P}}}\frac{\mathrm{d}}{\mathrm{dt}}\left({\mathrm{V}}_{\mathrm{F}}\mathrm{-}{\mathrm{V}}_{\mathrm{P}}\right)       
\end{equation}

As a result of rotation due to the velocity gradient, a force is applied to a single particle called the Lift force that can be calculated via Eq. (11) (Saffman \cite{saffman1965lift}):
\begin{equation}
F_L\mathrm{=}\frac{\mathrm{2}k{\nu }^{0.\mathrm{5}}{\rho }_fV_Pd_{ij}}{d_P{\left(d_{ij}d_{ji}\right)}^{0.\mathrm{25}}}\left(V_F\mathrm{-}V_P\right)     
\end{equation}

In which $d_{ij}$ is the deformation tensor and $k$ is considered to be $2.594$. The equation utilized to calculate lift force is suitable for fine particles. It should be said that Eq. (11) is practicable for submicron-sized particles (Sommerfeld \cite{crowe2011multiphase}).

\noindent Energy equation is written as:
\begin{equation}
\mathrm{\nabla }\mathrm{\bullet }\left({\rho }_fC_{P,f}V_FT_f\right)\mathrm{=}\mathrm{\nabla }\mathrm{\bullet }\left(k_f\mathrm{\nabla }T_f\right)\mathrm{+}V_PQ                
\end{equation}

In which $Q$ is defined by the following relation that is the heat flux transferred between nano-particles and fluid. 
\begin{equation}
Q\mathrm{=}hA_P\left(T_P\mathrm{-}T_f\right)   
\end{equation}

Where $A_P$ is defined as the particle surface area and $h$ may be attained as follows.
\begin{equation}
\mathrm{h=}\frac{{\mathrm{k}}_{\mathrm{f}}\mathrm{Nu}}{{\mathrm{d}}_{\mathrm{P}}}  
\end{equation}

In Eq. (14), $Nu$ is obtained from Ranz --Marshal [28] relation as follows.

\begin{equation}
\mathrm{Nu=2+0}.\mathrm{6}{\mathrm{Re}}^{\frac{\mathrm{1}}{\mathrm{2}}}_{\mathrm{d}}{\mathrm{Pr}}^{\frac{\mathrm{1}}{\mathrm{3}}}
\end{equation}

Considering a control volume around one particle and then applying energy conservation equation for the control volume leads to

\begin{equation}
{\mathrm{m}}_{\mathrm{P}}{\mathrm{C}}_{\mathrm{P},\mathrm{P}}\frac{\mathrm{d}{\mathrm{T}}_{\mathrm{P}}}{\mathrm{dt}}\mathrm{=Q}
\end{equation}

\noindent In the modeling of solid phase, one can write
\begin{equation}
{\mathrm{\nabla }}^{\mathrm{2}}\mathrm{T=0} 
\end{equation}

\subsection{The Thermophysical properties of nano-fluid}

\noindent Determination of the physical properties of nano-fluids have been the center attention of many researchers in the past decades and many researches were done in this regard. In the present study, the base fluid is water and the particles are aluminum oxide which are the most commonly used particles in nano-fluids. The properties of these particles are brought in Table 2 \cite{vajjha2010development}. As seen, the conductivity coefficient in nano-particles is much higher than pure water.

\subsection{ Boundary Conditions}

\noindent As shown in Fig. 1, a steady state and laminar fluid (pure fluid or nano-fluid) flow with uniform profile enters the channel at 25 ° C. Reynolds number is evaluated applying equation (18). In equation (18), properties are calculated at entrance temperature.

\begin{equation}
\mathrm{Re=}\frac{\mathrm{\rho }{\mathrm{u}}_{\mathrm{m}}{\mathrm{D}}_{\mathrm{h}}}{\mathrm{\mu }}   
\end{equation}

Where $D_h$ is the channel hydraulic diameter and is defined as below

\begin{equation}
{\mathrm{D}}_{\mathrm{h}}\mathrm{=}\frac{\mathrm{4}{\mathrm{H}}^{\mathrm{2}}\mathrm{-}\mathrm{\pi }{\mathrm{B}}^{\mathrm{2}}}{\mathrm{16H+4}\mathrm{\pi }\mathrm{B}}
\end{equation}

In equation (19), H and D are the channel height and rigid cylinder diameter, respectively.

\noindent A constant and uniform 200 ${\mathrm{W}}/{{\mathrm{m}}^{\mathrm{2}}}$ heat flux on the channel walls is selected as the thermal boundary condition. Then, fluid is discharged to the atmosphere. (see Fig. 1)
\subsection{ Numerical Solution}
\noindent Numerical methods are very applicable in the field of nano-fluid flow \cite{vajjha2010development,tehrani2017network,khodabandeh2017parametric,akbari2016numerical}. For discretization of the aforesaid nonlinear equations, the control volume method is used.  For estimation of the diffusion and convection terms, the upwind scheme of second order is used and the SIMPLE algorithm is employed for coupling the velocity and pressure fields.
\subsection{Grid Generation and Grid Sensitivity}
\noindent Grid generation is an important part of simulation, as it affects time, convergence, and results of the solution. Furthermore, regular gridding has better effects on the aforesaid parameters than irregular one dose. It is worth noting that grid should be fine enough near the walls so that one could investigate steep gradients of physical properties adjacent to the walls. These gradients occur perpendicular to the walls. Therefore, regular gridding with elements increment along the radius of the walls is used. In Fig. 2, a view of channel geometry gridding is depicted.

\noindent 

\noindent In order to check grid sensitivity, local Nusselt variations at the channel exit due to an increase in the grid elements is evaluated.

\noindent Water at 25 ° C and $Re=1000$ enters the channel shown in Fig 1, possessing the properties listed in Table 1. A constant $20000\ {W}/{m^2}$ heat flux is applied on the channel walls. Variations of Nusselt number with elements number at the channel exit is plotted in Fig. 3. By minifying grid, Nusselt number varies at the channel exit. As it can be observed, a grid having 1456446 elements is pertinent and no sensible change is seen minifying grid anymore. Elements number in five different cases is presented in Table 3.

\subsection{Validation}
\noindent After determining elements number, by applying nano-fluid properties equations and boundary conditions into the software, simulation is completed. For mass, energy, and momentum conservation equations, the minimum divergence criteria is considered to be${10}^{-5}$.

\noindent In order to verify the effect of nanofluids on heat transfer enhancement, an experimental study was done by Kim et al. \cite{kim2009convective}. For validating the results of the simulation, data presented in Ref. [35] are used. A laminar Al${}_{2}$O${}_{3}$-water with 3\% volume concentration is modeled in a 2-meter long cylinder having 4.57 mm of diameter under$2089.56\ {W}/{m^2}$ Constant heat flux with$Re=1460$. Results obtained are demonstrated in Fig. 4.

\noindent Considering presented figures in this section and good agreement with theoretical and experimental data, it is found that simulation is accurate enough and results are confirmed.

\section{Results and Discussion}

\subsection{Effects of Reynolds Number}
\noindent Fig. 5 represents axial variations of convective heat transfer coefficient and local Nusselt number of the base fluid versus Reynolds number. As Reynolds number increases, convective heat transfer and local Nusselt number grow due to the reduction in the boundary layer thickness.

\noindent In Fig. 6, axial variations of local convective heat transfer coefficient and local Nusselt number of Al${}_{2}$O${}_{3}$ nano-fluid is schemed against Reynolds number under 1\% volume concentration and nano-particles diameter of 25 nm. As stated for the base fluid, an increase in Reynolds number causes the convective heat transfer coefficient to increase.

\noindent For other concentrations, analogous graphs can be drawn. In all these graphs, variations of local convective heat transfer coefficient with Reynolds number is similar.

\noindent Effects of Reynolds number on the nano-fluid flow was investigated in local graphs. Now, by computing mean convective heat transfer coefficient, a better comparison can be performed. In Fig. 7, variations of mean convective heat transfer coefficient of Al${}_{2}$O${}_{3}$ nano-fluid versus Reynolds number is plotted for five different volume concentrations of nano-particles having 25 nm diameter. Mean convective heat transfer coefficient for water increases up to 181 percent, if Reynolds number changes from 100 to 1000, while for Al${}_{2}$O${}_{3}$ nano-fluid in 1\% and 5\% volume concentration, up to 184.4 and 199.6 percent growth is achieved.

\noindent Generally, by increasing Reynolds number, convective heat transfer for nano-fluid enhances as it did for the base fluid.
\subsection{Effects of Nano-Particles Concentration}
\noindent Another effective parameter on nano-fluid heat transfer enhancement is volume concentration of particles. Note that by changing concentration, two phenomena affect mean heat transfer from walls. On one side, adding nano-particles increases boundary layer thickness and reduces temperature gradient next to the walls. On the other side, nano-fluid heat transfer increases with concentration. The combined net effect is convective heat transfer coefficient enhancement. Also, nano-fluid density increases, if a rise in the volume fraction of particles occurs which leads to increase in momentum and convective heat transfer. Density gradient between particles and fluid temperature difference is the most important factors in naturally blended mixtures. Fluctuations in local density create a cavity adjacent to the fluid molecules that forces them to move and mix.

\noindent In Figs. 8 and 9, effects of volume concentration on heat transfer coefficient and local Nusselt number of the nano-fluid are depicted for Re=100, 500 and 1000. Results show that heat transfer coefficient and local Nusselt number of the nano-fluid in different Reynolds numbers increase with any rise in volume concentration. According to the relation of nano-fluid heat transfer with particles concentration increase, static and dynamic heat transfer of nano-fluid enhance which would increase heat transfer coefficient.

\noindent In Fig. 10, variations of convective heat transfer coefficient for Al${}_{2}$O${}_{3}$-water nano-fluid is plotted versus volume concentration under nano-particles diameter of 25 nm and Re=100, 250, 500, 750 and 1000. As it can be observed, increasing concentration at a constant Reynolds number leads to increase in convective heat transfer coefficient. For instance, in 5\% volume concentration and Re=100, convective heat transfer coefficient increases up to 26.47 percent relative to the base fluid. Also, for Re=500 and 1000, up to 33.13 and 34.71 percent increase can be obtained.

\subsection{Effects of Nano-Particles Size}

\noindent Another effective and basic parameter on convective heat transfer for nano-fluid is nano-particles size. According to the relation of heat transfer coefficient, one can infer that nano-particles size has negative effects on heat transfer and heat transfer coefficient. In Fig. 11, axial variations of convective heat transfer coefficient and local Nusselt number against size and volume concentration of nano-particles is demonstrated at Re=500. Results indicate that by increasing particles size at a constant volume concentration, local convective heat transfer coefficient decreases slightly. Variations of mean convective heat transfer coefficient for Al${}_{2}$O${}_{3}$-water nano-fluid versus size and volume concentration of nano-particles at Re=500 for water as base fluid is illustrated in Fig. 12. Alike local graphs, by decreasing particles size, nano-fluid heat transfer coefficient increases due to the rise in surface-to-volume ratio and properties improvement. For example, by changing particles diameter from 25 nm to 50 nm, mean convective heat transfer coefficient for Al${}_{2}$O${}_{3}$-water at 1\% and 5\% volume concentration reduces from 1.13 to 7.4 percent. It seems that at higher concentrations, nano-particles inside the base fluid show more reduction.

\subsection{Effects of Pressure Drop}
\noindent Increasing particles density and volume fraction lead to rising in momentum and convective heat transfer. Also, nano-fluid viscosity increases, if particles volume concentration enhances. Therefore, by increasing concentration, nano-fluid pressure drop increases relative to the base fluid. Fig. 13 displays variations of pressure drop for Al${}_{2}$O${}_{3}$-water versus Reynolds number at five nano-particles volume concentrations. Similar to the base fluid, rising Reynolds number at constant concentration, increases nano-fluid pressure drop. For instance, by changing Reynolds number from 250 to 750 at 1\% and 5\% volume concentration, pressure drop increases up to 240.7 and 228.3 percent, respectively. Also, slope of pressure drop experiences a growth with volume concentration.

\noindent Fig. 14 indicates that by increasing nano-particles volume concentration at constant Reynolds numbers, pressure drop rises. As an example, at Re=500 and 4\% volume concentration, pressure drop for nano-fluid is 1.88 times bigger than that of base fluid. Variations of nano-particles diameter do not have sensible influence on pressure drop.

\section{Conclusion}

\noindent A wide variety of processes involving heat and energy, have low efficiency due to low thermal conductivity of working fluid such as water, ethylene glycol, and engine oil. An advance method proposed to enhance heat transfer characteristic of fluid, is dispersion of ultrafine solid particles in a base fluid known as nano-fluids. In this paper, heat transfer and effects of Reynolds number, nano-particles concentration, nano-particles size, pressure drop and wavy sheet AL${}_{2}$O${}_{3}$ and nano-particles and water as the base fluid inside a square channel having an inner cylinder, with and without fin under constant heat flux condition were investigated. According to the research done, the following conclusions can be inferred. 
\begin{itemize}
	\item  As the thermal boundary layer grows, local Nusselt number decreases continuously.
	
	\item  Increasing mass flow and Reynolds number make a rise in heat transfer coefficient, base fluid, and nano-fluid Nusselt number. The more concentrated nano-fluid, the more increase in aforesaid parameters.
	
	\item  Adding nano-particles increases boundary layer thickness and reduces temperature gradient next to the walls. Also, nano-fluid heat transfer increases with concentration. The combined net effect is convective heat transfer coefficient enhancement. 
	
	\item  By decreasing particles size at a constant volume concentration, local convective heat transfer coefficient increases due to the rise in surface-to-volume ratio and properties improvement.
	
	\item  By increasing concentration, nano-fluid pressure drop increases relative to the base fluid. Similar to the base fluid, rising Reynolds number at constant concentration, increases nano-fluid pressure drop.
	
	\item  If Reynolds number rises, skin friction increases as well.
	
	\item  Flat separating sheet reduces convective heat transfer relative to the case where no sheet is used, while wavy sheet leads to convective heat transfer enhancement.
\end{itemize}

\noindent The extension of this paper for nanofluids, according to previous studies \cite{khodabandeh2016evaluation,tehrani2017cell,asadi2013novel,tehrani2017effect,tehrani2017revisiting,tehrani2017force,akbari2016investigation,kasaeian2017experimental,tehrani2017micromechanical,tehrani2017network,moghaddam2016metallic,skoracki2015additive}, affords engineers a good option for micro- and nano simulations. 
\newpage

\begin{tabular}{|p{0.5in}|p{2.0in}|p{0.5in}|p{2.0in}|} 
	\multicolumn{2}{|p{1in}|}{Notation} & $\mathrm{P}$ & Pressure, N/m2 \\ 
	$\mathrm{A}$ & Particle surface area, $m^{2} $ & Q & heat flux \\ 
	$\mathrm{B}$ & channel height & Re & Reynolds  \\ 
	${\mathrm{C}}_{\mathrm{c}}$ & Cunningham correction factor to Stokes' drag law & ${\mathrm{S}}_0$ & Spectral intensity basis \\ 
	${\mathrm{C}}_{\mathrm{p}}$ & Specific heat, J/kg K & ${\mathrm{S}}_{\mathrm{n,ij}}$ & Spectral intensity \\ 
	D & rigid cylinder diameter & $\mathrm{t}$ & Time (s) \\ 
	${\mathrm{d}}_{\mathrm{p}}$ & Particle diameter, nm & $\mathrm{T}$ &  Temperature, K \\ 
	${\mathrm{d}}_{ij}$ & Deformation tensor & v & Velocity, m/s \\  
	${\mathrm{F}}_{\mathrm{D}}$ & Drag force &  &  \\ 
	${\mathrm{F}}_{\mathrm{L}}$ & Lift force & \multicolumn{2}{|p{2.5in}|}{Greek symbols} \\  
	${\mathrm{F}}_{\mathrm{V}}$ & Virtual mass force & $\mu$ & Dynamic viscosity (${N\ s}/{m^2})$ \\ 
	${\mathrm{F}}_{\mathrm{G}}$ & Gravity force & ${\mathrm{\delta }}_{\mathrm{ij}}$ & Kronecker delta function \\ 
	${\mathrm{F}}_{\mathrm{P}}$ & Pressure gradient force & $\Delta $ & Difference \\  
	${\mathrm{F}}_{\mathrm{B}}$ & Brownian force & ${\mathrm{\zeta }}_{\mathrm{i}}$ & Zero-mean, unit-variance-independent Gaussian random number \\  
	$\mathrm{g}$ & Gravity acceleration, $m/s^{2} $ & $\mathrm{\lambda }$ & Molecular free path  \\  
	$\mathrm{h}$ & Convective heat transfer coefficient, $W/m^{2} K$ & $\mathrm{\upsilon }$ & Kinematic viscosity (${m^2}/{S})$ \\ 
	$\mathrm{k}$ & thermal conductivity for Fluid, W/m.K & $\mathrm{\rho }$ & Density\textbf{} \\  
	${\mathrm{k}}_{\mathrm{b}}$ & Boltzmann constant (=$1.3807\times 10^{23} J/k$) & \multicolumn{2}{|p{2.5in}|}{Subscripts} \\ 
	$\mathrm{L}$ & axial length, m & f & Fluid  \\ 
	$\mathrm{Nu}$ & Peripherally average Nusselt number & p & Particle  \\ 
\end{tabular}

\newpage

\begin{table}[h!]
\centering
\caption{Dimension of the channel under consideration}
\begin{tabular}{c c c c} \hline 
	L (m) & A (mm) & B (mm) & D (mm) \\ \hline 
	3 & 50 & 50 & 25 \\ \hline 
\end{tabular}

\label{table:1}
\end{table}

\begin{table}[h!]
	\centering
	\caption{the properties of ${Al}_2O_3$ [31]}
\begin{tabular}{c c} \hline 
	Value & Property \\ 
	3600\textit{} & Density ($\mathrm{kg/}{\mathrm{m}}^{\mathrm{3}}$) \\ 
	765\textit{} & Specific heat$\mathrm{(}\frac{\mathrm{J}}{\mathrm{kg\ K}}\mathrm{)}$ \\ 
	36\textit{} & Thermal Conductivity $\mathrm{(}\frac{\mathrm{w}}{\mathrm{mk}}\mathrm{)}$ \\ 
	50 & Diameter of particle $\mathrm{(nm)}$ \\ \hline 
\end{tabular}

\label{table:2}
\end{table}

\begin{table}[h!]
	\centering
	\caption{Cell number in cylinder for case 3}
\begin{tabular}{c c} \hline 
	Grid & Cell Number \\ \hline
	Grid 1 & 270600 \\ 
	Grid 2 & 416000 \\
	Grid 3 & 960000 \\ 
	Main Grid & 1456446 \\ 
	Grid 5 & 2949442 \\ \hline 
\end{tabular}

\label{table:3}
\end{table}	
	
\newpage

\begin{figure}[h]
	\centering
	\includegraphics[width=1\textwidth]{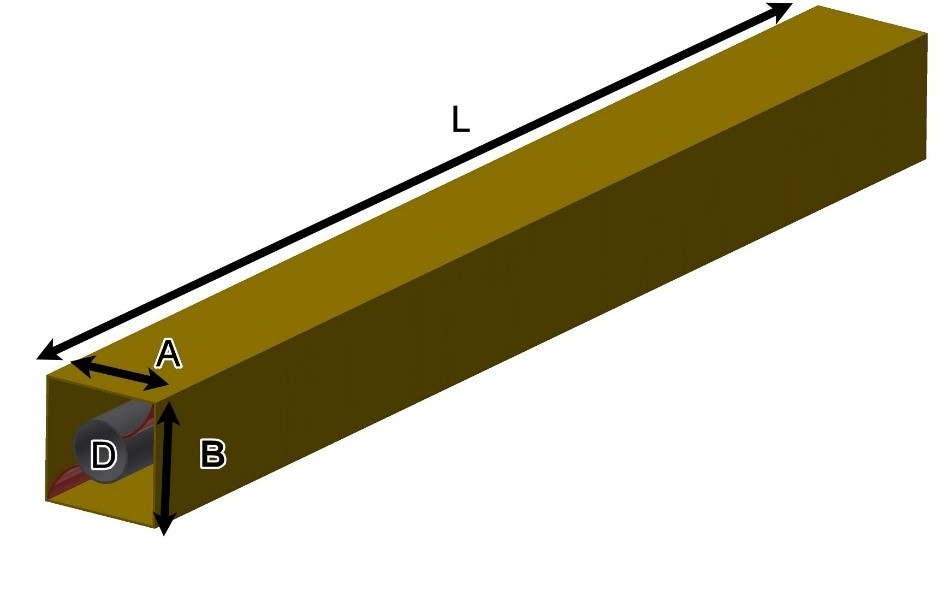}
	\caption{Cases study geometry, Channel with an Inner Cylinder and without fin}
\end{figure}
\newpage
\begin{figure}[h]
	\centering
	\includegraphics[width=1\textwidth]{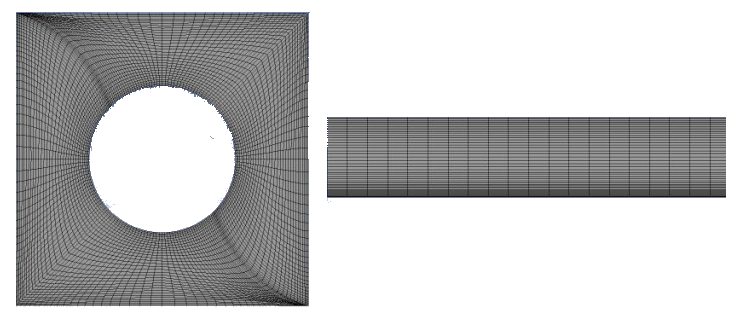}
	\caption{channel griding, isometric view}
\end{figure}
\newpage
\begin{figure}[h]
	\centering
	\includegraphics[width=1\textwidth]{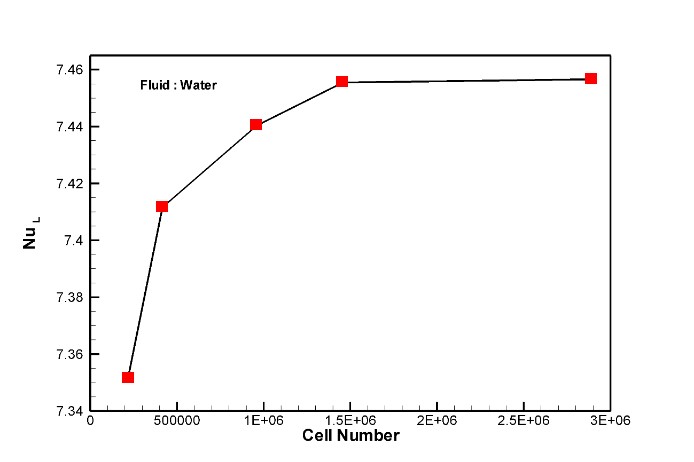}
	\caption{variations of Nusselt number at channel exit under constant heat flux condition}
\end{figure}
\newpage
\begin{figure}[h]
	\centering
	\includegraphics[width=1\textwidth]{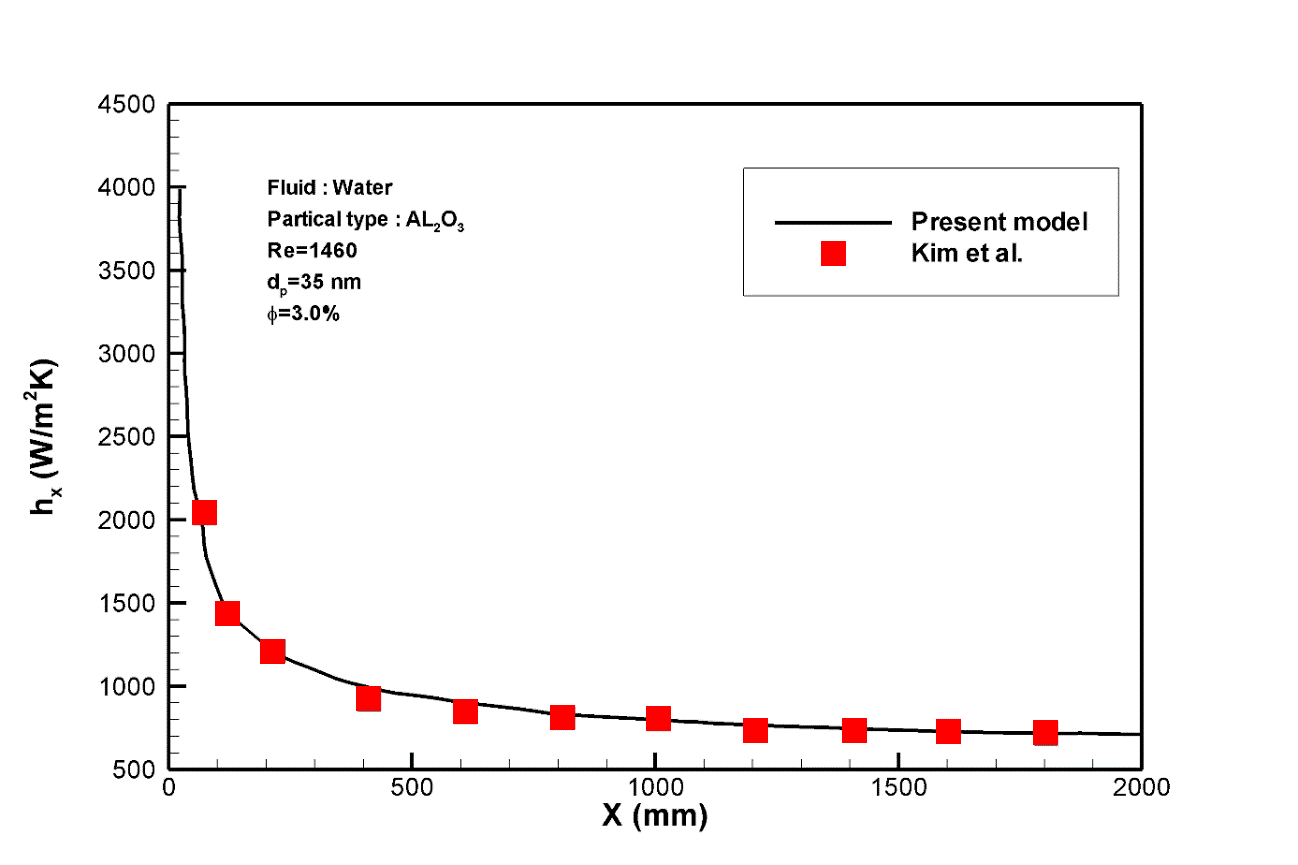}
	\caption{ axial variations of local convective heat transfer coefficient for Al${}_{2}$O${}_{3}$ nano-fluid in 3\% volume concentration, nano-particles of 35 nm and $Re=1460$.  }
\end{figure}
\newpage
\begin{figure}[h]
	\centering
	\includegraphics[width=1\textwidth]{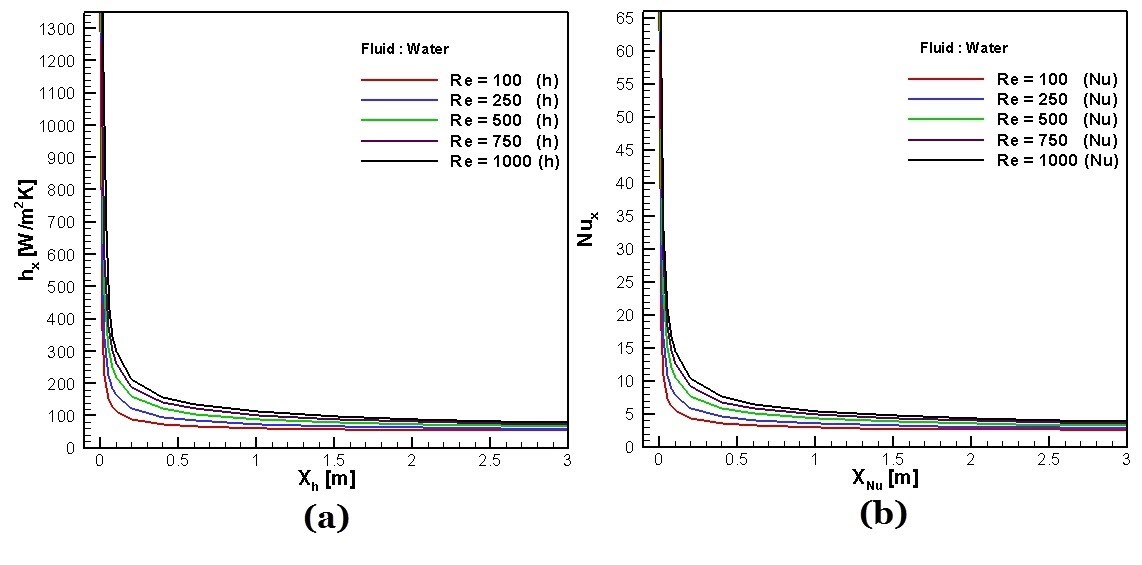}
	\caption{axial variations of (a) convective heat transfer coefficient and (b) local Nusselt number of the base fluid with respect to the Reynolds number.  }
\end{figure}
\newpage
\begin{figure}[h]
	\centering
	\includegraphics[width=1\textwidth]{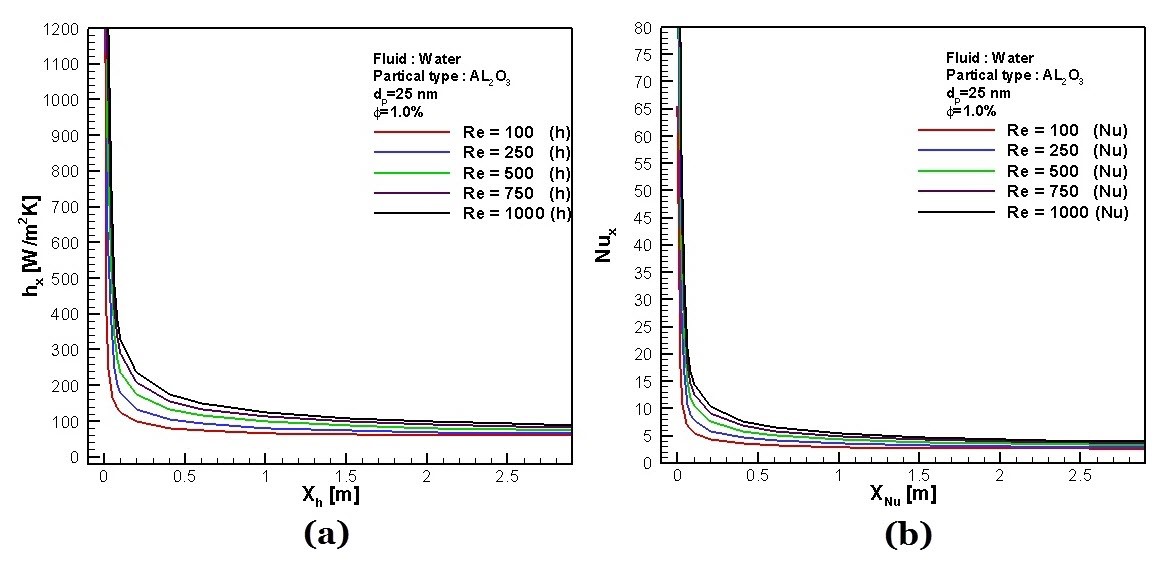}
	\caption{axial variations of (a) local convective heat transfer coefficient and (b) local Nusselt number of Al${}_{2}$O${}_{3}$ nano-fluid versus Reynolds number under 1\% volume concentration and nano-particles diameter of 25 nm.  }
\end{figure}
\newpage
\begin{figure}[h]
	\centering
	\includegraphics[width=1\textwidth]{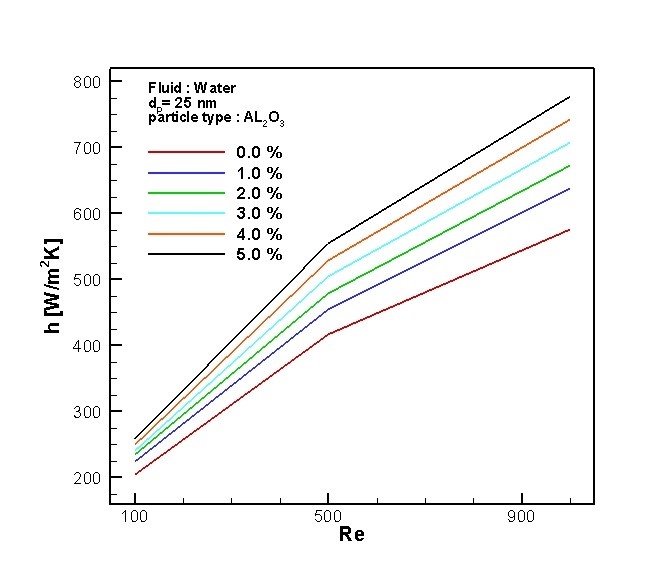}
	\caption{variations of mean convective heat transfer coefficient of Al${}_{2}$O${}_{3}$ nano-fluid versus Reynolds number for five different volume concentrations of nano-particles having 25 nm diameter.  }
\end{figure}
\newpage
\begin{figure}[h]
	\centering
	\includegraphics[width=1\textwidth]{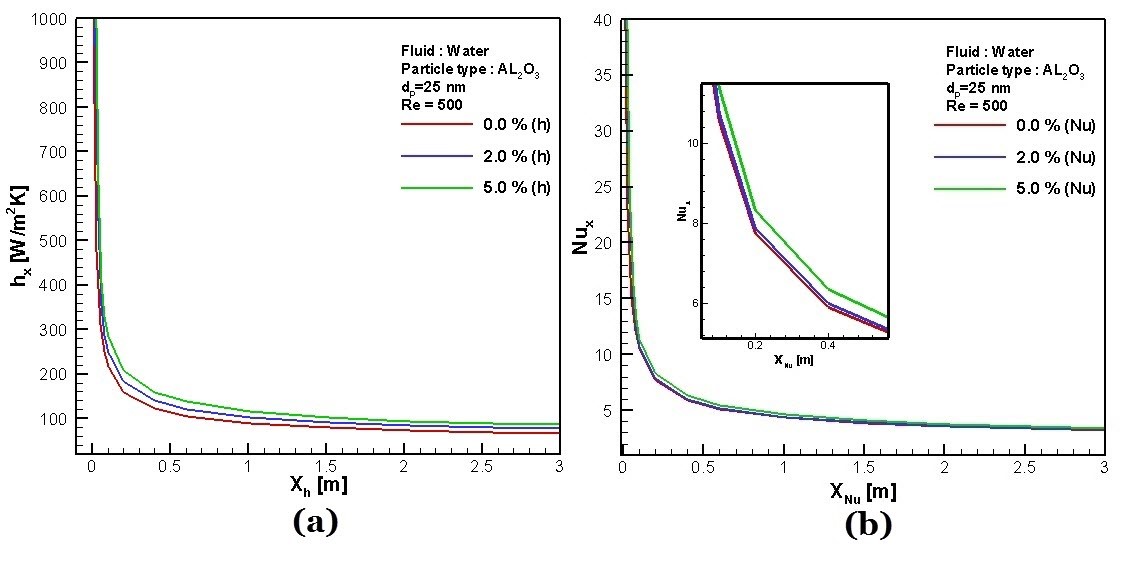}
	\caption{ axial variations of (a) local convective heat transfer coefficient and (b) local Nusselt number of Al${}_{2}$O${}_{3}$-water nano-fluid versus nano-particles volume concentration at Re=500.  }
\end{figure}
\newpage
\begin{figure}[h]
	\centering
	\includegraphics[width=1\textwidth]{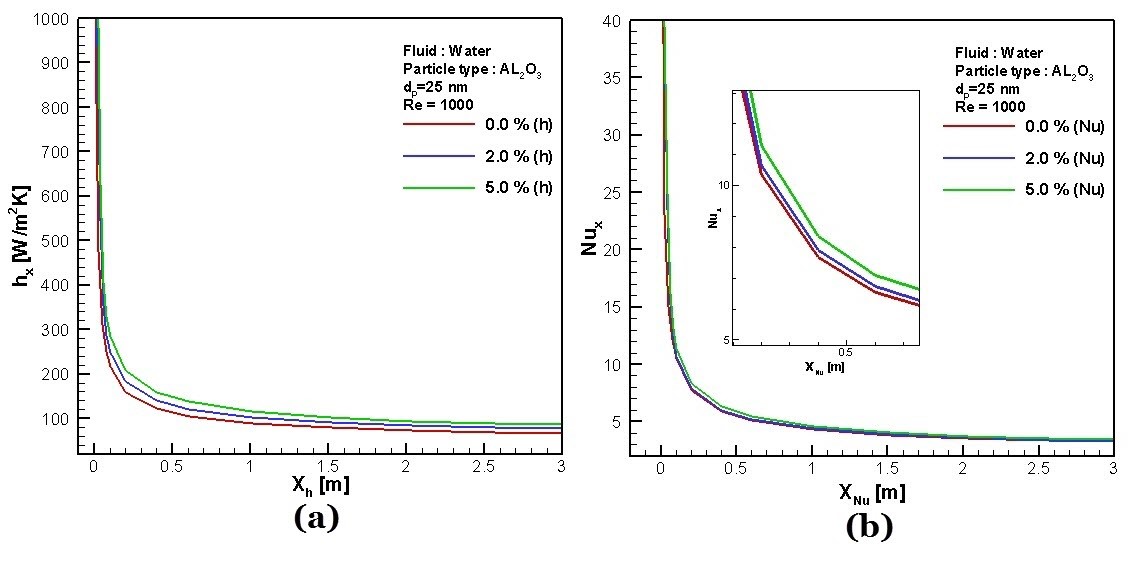}
	\caption{axial variations of (a) local convective heat transfer coefficient and (b) local Nusselt number of Al${}_{2}$O${}_{3}$-water nano-fluid versus nano-particles volume concentration at Re=1000.  }
\end{figure}
\newpage
\begin{figure}[h]
	\centering
	\includegraphics[width=1\textwidth]{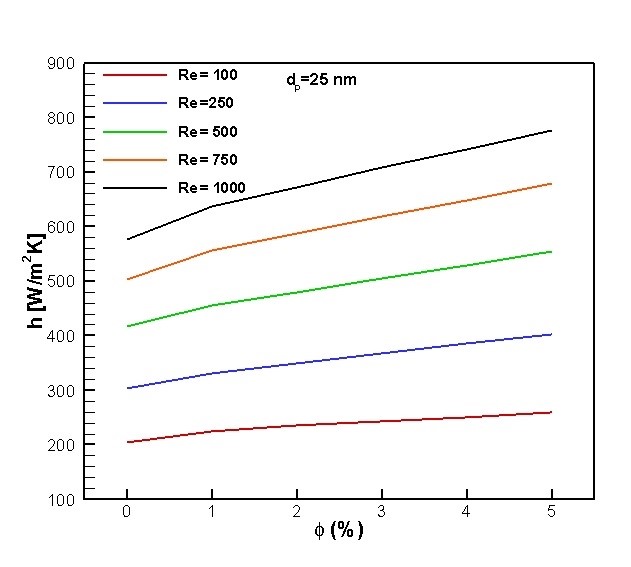}
	\caption{variations of convective heat transfer coefficient for Al${}_{2}$O${}_{3}$-water nano-fluid versus volume concentration under nano-particles diameter of 25 nm and Re=100, 250, 500, 750 and 1000.  }
\end{figure}
\newpage
\begin{figure}[h]
	\centering
	\includegraphics[width=1\textwidth]{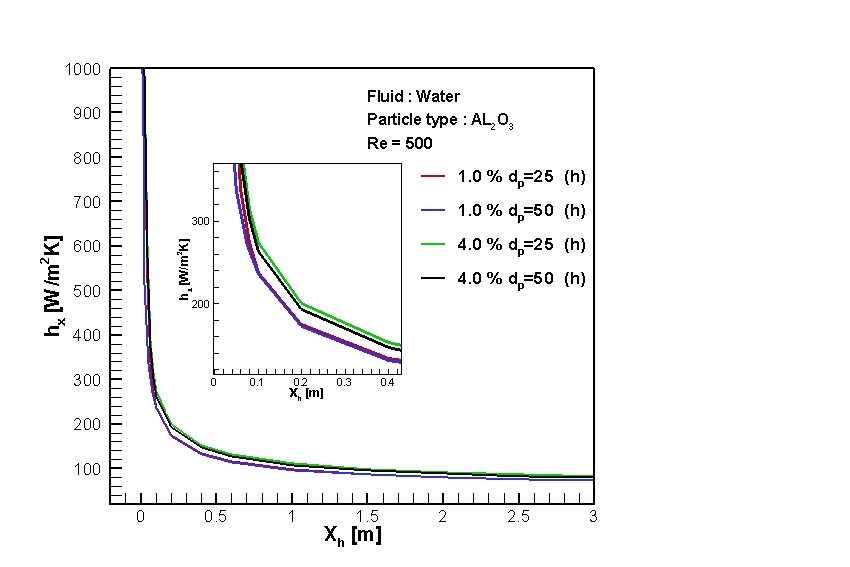}
	\caption{axial variations of convective heat transfer coefficient  against size and volume concentration of nano-particles at Re=500.  }
\end{figure}
\newpage
\begin{figure}[h]
	\centering
	\includegraphics[width=1\textwidth]{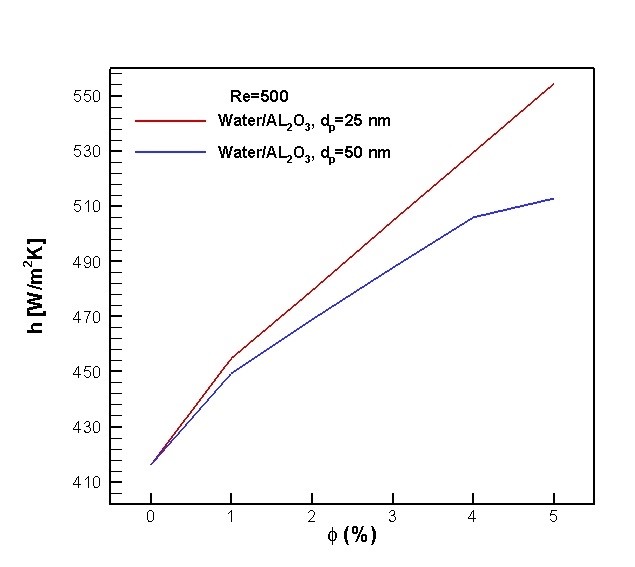}
	\caption{variations of convective heat transfer coefficient and mean Nusselt number against size and volume concentration of nano-particles at Re=500.  }
\end{figure}	
	
\newpage	
	\begin{figure}[h]
		\centering
		\includegraphics[width=1\textwidth]{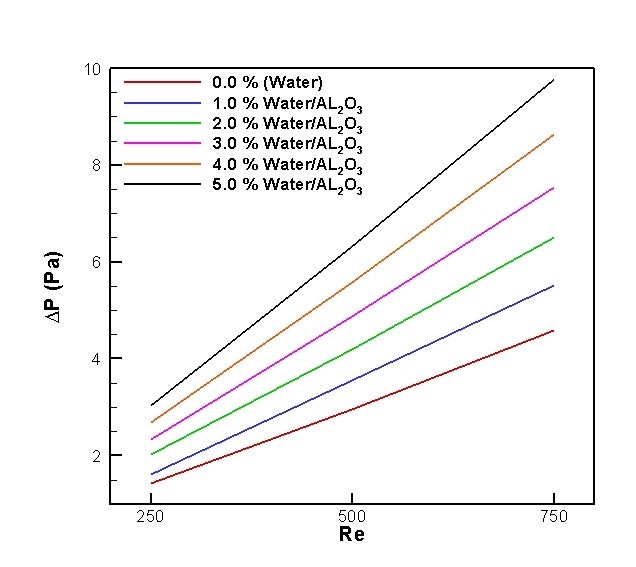}
	\caption{variations of Al${}_{2}$O${}_{3}$-water pressure drop versus Reynolds number at five nano-particles volume concentrations having diameter of 25 nm.  }
	\end{figure}	
\newpage	
	\begin{figure}[h]
		\centering
		\includegraphics[width=1\textwidth]{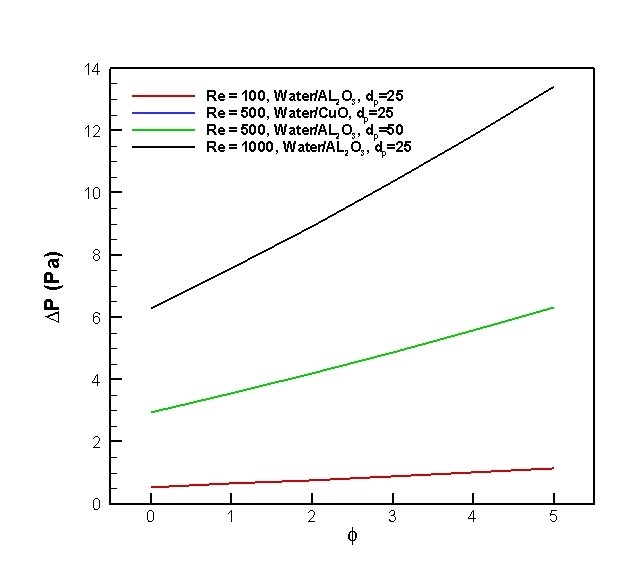}
	\caption{variations of pressure drop for Al${}_{2}$O${}_{3}$-water against volume concentration at two different diameters and Reynolds numbers.  }
	\end{figure}	
\newpage	
	\begin{figure}[h]
		\centering
		\includegraphics[width=1\textwidth]{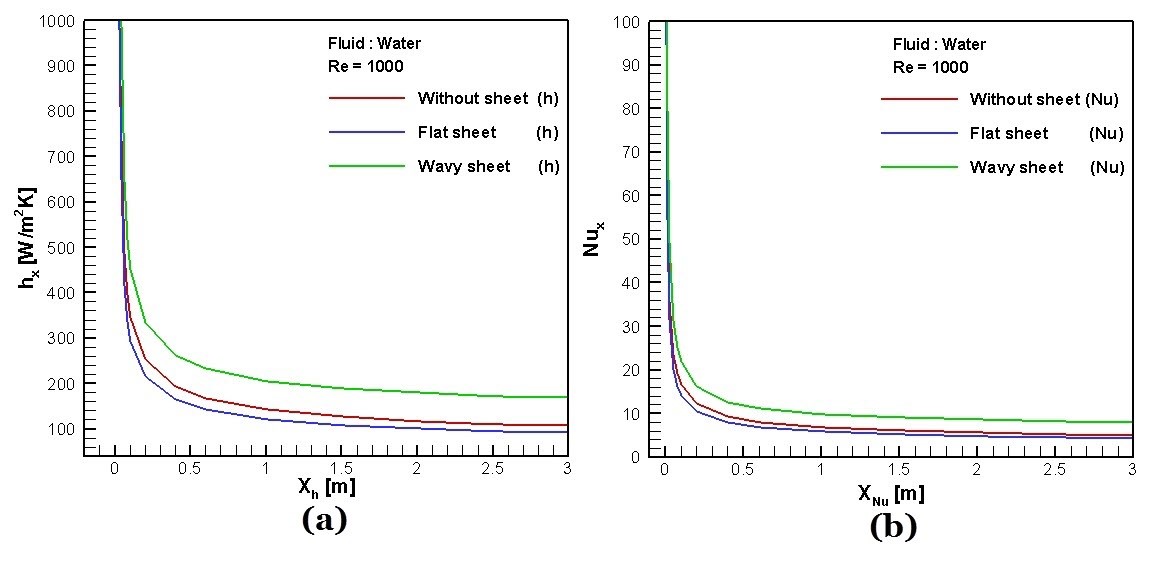}
	\caption{variations of (a) convective heat transfer coefficient and (b) Nusselt number against separating sheet shape at Re=1000.  }
	\end{figure}	
\newpage

		\bibliography{ERF}
	
\end{document}